\documentclass[aps,prl,twocolumn,superscriptaddress,a4paper]{revtex4}
\usepackage{amsmath,amssymb,graphicx,dcolumn,bm,url}
\usepackage[colorlinks=true,linkcolor=red,citecolor=blue]{hyperref}
\usepackage{mathptmx}
\usepackage{amsfonts}

\begin{document}
\title{Spectral Shearing of Quantum Light Pulses by Electro-Optic Phase Modulation}

\author{Laura J. Wright}
\affiliation{Clarendon Laboratory, University of Oxford, Parks Road, Oxford, OX1 3PU, UK}
\author{Micha{\l} Karpi{\'n}ski}
\email{mkarp@fuw.edu.pl}
\affiliation{Clarendon Laboratory, University of Oxford, Parks Road, Oxford, OX1 3PU, UK}
\affiliation{Faculty of Physics, University of Warsaw, Pasteura 5, 02-093 Warszawa, Poland}
\author{Christoph S{\"o}ller}
\affiliation{Clarendon Laboratory, University of Oxford, Parks Road, Oxford, OX1 3PU, UK}
\author{Brian J. Smith}
\email{bjsmith@uoregon.edu}
\affiliation{Clarendon Laboratory, University of Oxford, Parks Road, Oxford, OX1 3PU, UK}
\affiliation{Department of Physics and Oregon Center for Optical, Molecular, and Quantum Science, University of Oregon, Eugene, Oregon 97403, USA}
\begin{abstract}
Frequency conversion of nonclassical light enables robust encoding of quantum information based upon spectral multiplexing that is particularly well-suited to integrated-optics platforms. Here we present an intrinsically deterministic linear-optics approach to spectral shearing of quantum light pulses and show it preserves the wave-packet coherence and quantum nature of light. The technique is based upon an electro-optic Doppler shift to implement frequency shear of heralded single-photon wave packets by $\pm 200$~GHz, which can be scaled to an arbitrary shift. These results demonstrate a reconfigurable method to controlling the spectral-temporal mode structure of quantum light that could achieve unitary operation.
\end{abstract}

\maketitle

The frequency of a single light quantum, or photon, is a key physical property of individual excitations of the quantized electromagnetic field \cite{dirac1927}, which were introduced to describe the photoelectric effect \cite{einstein1905}. Frequency is a mode characteristic, just as polarization, transverse-spatial amplitude, and direction of propagation define the modes of electromagnetic radiation. Thus frequency can be transformed using linear-optical elements in much the same way lenses transform transverse-spatial modes and wave plates manipulate polarization modes. Frequency is not an immutable property of photons---it can be coherently and deterministically modified. For example, retroreflection from a moving mirror results in a frequency shift due to the Doppler effect \cite{einstein1905_2,ives1940}. The various independent degrees of freedom that comprise the modes of light can be used to encode information in the electromagnetic field, namely position-momentum, time-frequency, and polarization. Information-technology applications require precise means for manipulation and measurement of light in the encoding degree of freedom. Many preliminary demonstrations of quantum optical technologies have utilized polarization, path or transverse-spatial mode encoding. These degrees of freedom are limited to relatively few quantum bits that can be practically addressed per photon within an integrated-optics platform, in which high-stability, low-loss multiphoton interference, necessary for optical quantum technologies, can occur. Recently, the time-frequency (TF) mode structure of light has come to the fore in quantum photonics as an ideal means of quantum information encoding for integrated optical quantum technologies \cite{kielpinski11, hayat12, donohue14, motes14, sinclair14, brecht15, reimer16}. 

Essential to both quantum and classical technologies based upon TF mode encoding is the ability to control the pulse-mode structure of light---where the central frequency and arrival time play prominent roles. In the classical domain the primary methods to control an optical pulse are based upon direct modification of the wave packet by amplifying and filtering different frequency and time components \cite{dubietis92,weiner2000}. This approach to pulse shaping is incompatible with quantum states of light owing to noise and signal degradation arising from amplification and loss, resulting in destruction of the fragile quantum coherences between different photon-number components of the state. To preserve the quantum state of light when modifying the mode structure of optical pulses thus requires unitary, i.e., phase-only, evolution in principle. In practice, unitary state evolution is challenging to achieve due to technical losses, e.g., loss arising from transverse spatial mode mismatch between integrated optical components. However, deterministic control of optical pulsed modes, in which each pulse successfully passing through a device is precisely modified, can be attained by application of spectral and temporal phase.

A range of approaches to manipulate the TF state of quantum light pulses based upon nonlinear optical techniques have been demonstrated \cite{huang92,tanzilli05,takesue08,mcguinness10,lavoie13}. Here manipulation of the pulse mode structure of a weak quantum optical signal is implemented by nonlinear optical interaction with a strong classical control field mediated by a nonlinear medium. Although highly efficient operation has been demonstrated for these processes \cite{ates12, zaske12, vollmer14}, achieving fully deterministic operation with low noise can be challenging to realize due not only to additional nonlinear processes occurring simultaneously, but also to more fundamental limitations \cite{christ13,quesada16}. Recently, deterministic spectral shear based on nonlinear cross-phase modulation with a strong pump beam has been demonstrated \cite{matsuda16}. Such nonlinear approaches often require carefully tailored phase- and group-velocity-matching conditions and experimental arrangements that are difficult to apply across a broad spectral range and reconfigure. 

Here we present a technique to achieve deterministic spectral shearing of broadband quantum pulses by application of a linear temporal phase to the pulses using electro-optic phase modulation. We experimentally demonstrate $\pm 200$~GHz spectral shear of heralded single photons with $435$~GHz bandwidth. Two-photon interference between a tunable single-photon reference pulse and the spectrally sheared single photons demonstrates preservation of the pulsed wave-packet coherence. The nonclassical nature of the single-photon state, characterized by the heralded degree of second-order coherence $g_{\mathrm{h}}^{(2)}(0)$, is also shown to remain unchanged by the frequency shear operation. Electro-optic phase modulation enables high-fidelity operations on both continuous-wave light at the single-photon level \cite{merolla, olislager12, xing14}, and pulsed quantum light as we demonstrate here. The approach is compatible with pulses over a large range of central wavelengths, bandwidths and pulse shapes. The method can be applied repeatedly without modification of the experimental setup, allowing multiple shearing operations to be performed in sequence. The phase imprinted on the pulse is readily reconfigurable by changing the driving voltage applied to the modulator, allowing feed-forward control. The spectral shearing operation is intrinsically deterministic, paving a feasible path to unitary TF mode manipulation.

The electromagnetic field mode occupied by a single light quantum can be viewed as the single-photon wave function \cite{sipe95, birula94, smith07}. In the paraxial regime a beam of light has transverse-spatial, polarization and longitudinal time-frequency mode structure. Here we focus on the pulsed spectral-temporal modes, which are solutions to the wave equation in the slowly varying envelope approximation \cite{diels06, brecht15}. We refer to these as time-frequency (TF) modes, which can be expressed in the following complex form
\begin{equation}
\psi(t) = \left | \psi(t) \right| \exp \left\{i\left[\phi(t) - \omega_0 t \right] \right\},
\label{eq:1}
\end{equation}
\noindent where $\left | \psi(t) \right|$ is the amplitude of the slowly varying temporal envelope, $\phi(t)$ is the corresponding temporal phase, $\omega_0$ is the central angular frequency of the pulse, and $t$ is time in the reference frame of the pulse. Deterministic spectral shear by angular frequency, $\Omega$, can be realized by applying linear temporal phase across a pulse, $\phi_{\rm{ext}}(t) = - \Omega t$. In this case the initial normalized pulse mode amplitude, $\psi(t)$, is mapped to $\psi(t) \exp \left[-i \Omega t \right ]$. That linear temporal phase results in a spectral shear is a direct consequence of the Fourier shift property and Fourier relation between the temporal $\psi(t)$, and spectral $\tilde{\psi}(\omega)$, amplitudes.

\begin{figure} 
\includegraphics[width=8.6cm]{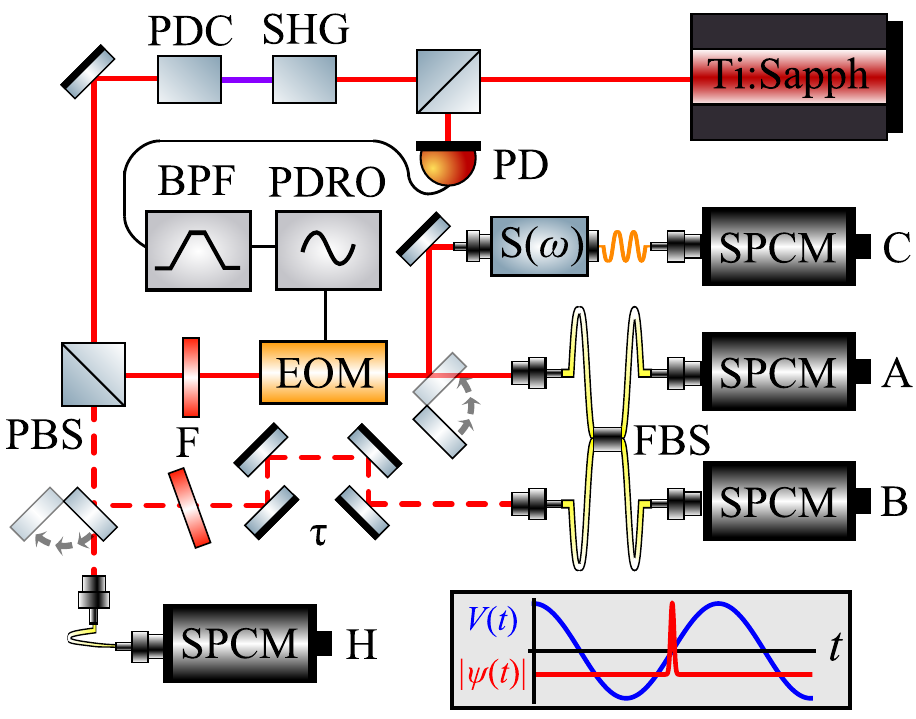}
\caption{Experimental setup for single-photon spectral shearing by electro-optic phase modulation. Photon pairs produced by parametric down-conversion (PDC), pumped by second-harmonic generation (SHG) of a pulse train from a titanium sapphire laser (Ti:Sapph), are separated at a polarizing beam splitter (PBS). The herald photon (dashed lines) can be directed to a single-photon counting module (SPCM H) to trigger the single-photon spectrometer, S$(\omega)$, and $g_{\mathrm{h}}^{(2)}(0)$ measurements at a fiber beam splitter (FBS), or directed to a FBS after passing through an interference filter (F) which can be tuned and adjustable time delay ($\tau$) to undergo two-photon interference with the signal photon (solid line). The EOM is driven by a sinusoidal rf field that is locked to a weak sample of laser pulse train monitored by a photodiode (PD). Inset: Schematic illustrating a linear temporal phase applied to a pulsed mode of amplitude $|\psi(t)|$ (red) by synchronization of sinusoidal EOM driving voltage $V(t)$ (blue) with the optical pulse train.} 
\label{fig:procedure}
\end{figure}

Our approach to linear temporal phase modulation of optical pulses by electro-optic phase modulation is depicted in the inset of Fig.\ 1. A short optical pulse undergoes a time-varying phase by passing through an electro-optic modulator (EOM) driven by a sinusoidal radio-frequency (rf) voltage signal, $V(t) = V_0 \sin(2\pi \nu t + \phi_0)$, that is synchronized with the optical pulse. Here $V_0$ is the driving voltage amplitude, $\nu$ the rf signal frequency, and $\phi_0$ is a controllable phase offset of the rf signal with respect to the optical pulse. Locking the linear region of the rf signal with an optical pulse results in a linear temporal phase applied to the pulse mode, provided the pulse temporal duration, $\tau_p$, is much less than the rf period $T=1/\nu$ \cite{duguay66}. That the optical pulse duration is shorter than the period of the rf driving field distinguishes the present approach from previous works where light with coherence time significantly longer than the rf modulation period was subjected to electro-optic phase modulation \cite{merolla,olislager12,xing14}. Constraining the pulse duration to be shorter than the sinusoidal rf period ensures that only a linear phase modulation is applied to the pulse, and defines the temporal aperture of the spectral shear operation. The magnitude and sign of the shift depend on the the slope of the phase modulation. Expanding about the linear region of the sinusoidally varying voltage signal gives an applied shear of $\Omega \approx \pm 2 \pi^2 (V_0/V_\pi )\nu$, where $V_\pi$ is the voltage required to achieve a $\pi$ phase shift with the modulator. The resulting shear varies linearly with the amplitude and frequency of the applied voltage, both of which can be readily modified to enable real-time feedback.

To demonstrate spectral shear of quantum light by applying linear temporal phase with an EOM we generate heralded single photons in temporally short optical pulses as depicted in Fig. 1. The heralded single photons are produced by collinear type-II spontaneous parametric down-conversion (SPDC) in potassium dihydrogen phosphate (KDP) \cite{mosley08}. The KDP crystal is pumped by a frequency-doubled Ti:sapphire laser oscillator operating at $80$~MHz repetition rate with $831.5$~nm central wavelength and $10$~nm full-width at half-maximum (FWHM) intensity bandwidth. The orthogonally polarized degenerate signal and idler fields are separated at a polarizing beam splitter. The heralded photon is spectrally filtered to a bandwidth of $\Delta \lambda = 1$~nm FWHM, to ensure that group-delay dispersion introduced by the $2$-m-long EOM fiber pigtail and the EOM itself does not lead to significant stretching of the single-photon temporal envelope and thus the pulse remains within the linear phase modulation region. The signal and idler modes are then coupled into polarization-maintaining single-mode fibers (PM-SMF). The idler mode is directed to a heralding single-photon counting module (SPCM), and the signal mode is directed to the EOM. 

The linear temporal phase modulation of the single-photon wave packet is realized by phase locking the EOM rf driving field with the laser pulse train from which the single-photon pulses are derived. A fast photodiode (PD, Thorlabs DET10A) monitors a weak sample of the laser pulse train, the output of which is passed through a narrow ($80$ kHz) bandpass filter (BPF, ASPA B80-3/T-6C) providing a timing reference for the sinusoidal field driving the EOM. A $\nu = 40$~GHz rf voltage is generated by a phase-locked dielectric resonator oscillator (PDRO, Herley-CTI PDRO-4000), which comprises a dielectric resonator oscillator locked with the $500$th harmonic of the filtered PD signal \cite{dorrer10}. Fine-tuning of the amplitude and phase offset of the rf signal from the PDRO is achieved using a variable attenuator (ATM, AT40A-3637-C40AV-06) and rf delay line (ATM, P28K-413) (not shown) to enable tunable control of the phase modulation applied to the optical pulses. The $40$~GHz signal is amplified to approximately $2$~W with a fast amplifier (Quinstar, QPN-40003330-A0) whose output was connected directly to the EOM (EOSpace PM-AV5-40-PFU-PFU-830-SRF1W).

Three key measurements before and after the spectral shear operation assess the performance of our technique---direct spectral characterization, heralded degree of second-order coherence, and two-photon interference. First, to demonstrate the operation does indeed perform deterministic spectral shear, the heralded spectrum of single photons is obtained. This measurement is implemented by scanning a multimode fiber ($50~\mu$m core diameter) in the image plane of a $300$-mm-focal-length spectrograph with a $1200$~lines/mm grating, the output of which is coupled to a SPCM as depicted in Fig. 1. Monitoring coincidence counts between the spectrometer output and the herald detector with a custom field-programmable gate array (FPGA) provides the heralded spectral intensity of the single photons \cite{kim05, mosley08}. By locking the positive (negative) slope of the sinusoidal rf driving signal to the heralded single-photon pulses we achieve deterministic spectral shear of $\Omega/2\pi = \pm 200$ GHz, as shown in Fig.\ 2, determined from Gaussian fit parameters. The data have been normalized to the same total number of counts per data set to compensate for drift in the single-photon generation rate. Independent measurements confirm that transmission through the EOM does not depend on the applied voltage. The demonstrated spectral shear is a significant fraction of the pulse bandwidth and can be increased by applying a stronger rf driving voltage \cite{farias05}, cascading many modulators, or using modulators with lower $V_{\pi}$ \cite{jouane14, zhang14}. These results directly show deterministic active modification of the spectral intensity profile of a nonclassical pulse through application of a precise temporal phase.
\begin{figure} 
\centering
\includegraphics[width=7cm]{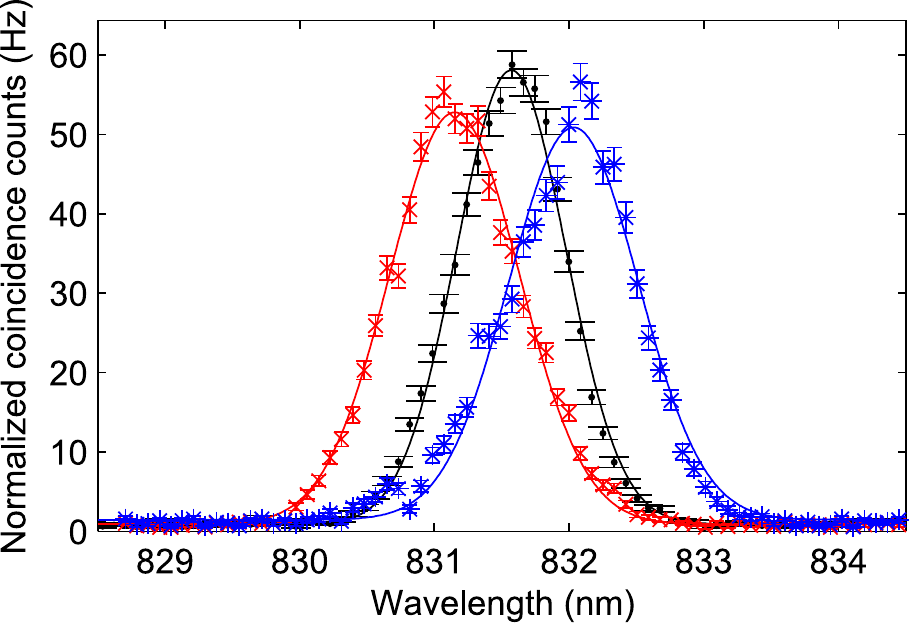}
\caption{Heralded single-photon spectra (data points) and Gaussian fits (solid lines) for the original pulse (black), positive (blue) and negative (red) linear temporal phase resulting in a spectral shear of $ \Omega/2\pi = \pm 200$ GHz. Uncertainties in coincidence counts are calculated assuming Poisson statistics and the uncertainty in wavelength is below the symbol size.} 
\label{fig:spectra}
\end{figure}

\begin{figure} 
\centering
\includegraphics[width=7cm]{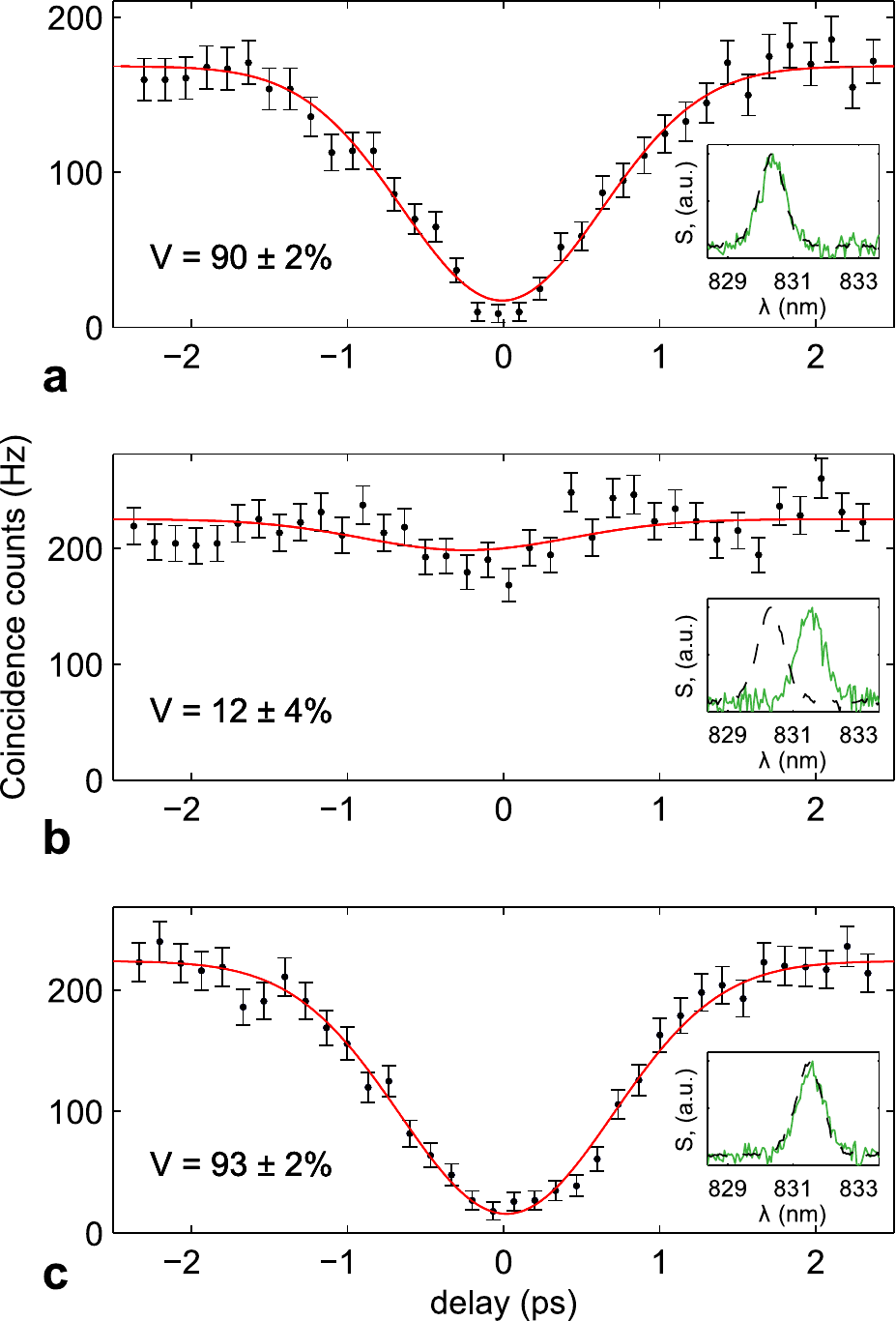}
\caption{Two-photon interference between spectrally sheared single-photon wave packets and a reference single photon derived from the idler mode of the SPDC. Coincidence counts between detectors $A$ and $B$ after subtracting background events obtained by blocking the signal and reference paths are plotted against delay $\tau$ between the signal and reference. The three plots correspond to different spectral shear and reference pulses (a) maximum blueshifted single photon and matched reference, (b) maximum redshifted single photon and reference matched to blueshifted photon, and (c) maximum redshifted single photon and matched reference. Error bars assume Poisson count statistics. We report two-photon interference visibilities obtained from inverted Gaussians (solid red) fitted to the data. Insets show spectra for the reference (dashed black) and sheared (solid green) single photons.} 
\label{fig:dips}
\end{figure}

The ideal spectral shearing operation should modify only the mode structure of the input light and not change the photon-number distribution of the state. This implies that the process should not add or remove photons from the state. Spectral shearing by electro-optic phase modulation is, in principle, a lossless unitary process. However, due to the transverse-spatial-mode mismatch between the optical fibers and the optical waveguide within the EOM, transmission through the current implementation is $0.5$, which can be improved with different input-output coupling. To monitor the nonclassical nature of the single-photon state and ensure no additional photon-number noise is produced during the process, measurement of the heralded degree of second-order coherence, $g_{\rm{h}}^{(2)}(0)$, is performed \cite{beck07}. The signal mode is split at a $50$:$50$ fiber coupler with both outputs monitored by SPCMs, labeled $A$ and $B$, as depicted in Fig. 1. Single-detector, $N_j$, two-fold, $N_{jk}$, and triple, $N_{jkl}$, coincidence count events, $j,k,l \in \{A, B, H\}$, are recorded to determine $g_{\rm{h}}^{(2)}(0) = N_{ABH}N_{H}/\left(N_{AH}N_{BH}\right)$. Prior to application of the phase modulation the source exhibited $g_{\rm{h}} ^{(2)}(0) = 0.038 \pm 0.001$ and after modulation it was found to be $g_{\rm{h}}^{(2)}(0) = 0.040 \pm 0.001$, where errors are calculated assuming Poisson count statistics. The measurements were performed over a period of approximately $15$~h each to attain sufficient threefold coincidence count events. These results demonstrate that the deterministic spectral shearing operation does not introduce measurable amplitude noise through the addition of photons and thus preserves the fragile quantum signal. 

An essential specification for deterministic mode manipulation of quantum light is the coherence of the process. This means that the process does not introduce phase noise between different input-output modes. In the case of pulsed wave-packet modes this implies that the phase relationship between different frequency components must be preserved. To demonstrate our spectral-shear operation preserves this wave-packet coherence, two-photon Hong-Ou-Mandel interference between the spectrally sheared single photon and a reference single-photon pulse is performed. The visibility of the two-photon interference measurement bounds the overlap of the two input photon modes \cite{hendrych03, takesue08, ates12}. Using the filtered idler single-photon wave packet as a reference and tuning the interference filter to optimize the spectral overlap with the $200$~GHz blueshifted signal photon, two-photon interference with $(90 \pm 2 )\%$ background-subtracted visibility is achieved, as shown in Fig.\ 3(a). Then tuning to the maximum redshifted spectral shear, $-200$~GHz, of the signal photon while leaving the reference unchanged results in two-photon interference with $(12 \pm 4)\%$ visibility, Fig.\ 3(b), which is consistent with the spectral mode overlap including the spectral phase introduced by the interference filters. Note that this result does not display oscillations in the coincidence signal as a function of the time delay, as found in the work of Ou and Mandel \cite{ou88}. The key difference resides in the fact that we do not perform spectrally resolved measurements, which in Ref.\ \cite{ou88} is implemented by placing interference filters in front of the detectors, but after the beam splitter. In our experiment the filters are placed before the beam splitter. Finally, by tilting the interference filter in the reference idler beam to optimize the spectral overlap with the redshifted signal photon results in $(93 \pm 2)\%$ two-photon interference visibility, Fig.\ 3(c). We verified that deviation from unit visibility arises from the photon pair source itself by measuring two-photon interference visibility between unmodified signal photons and an appropriately tuned reference. The maximum measured visibility of $(90 \pm 3) \%$ sets an upper bound on visibilities achievable in our experiment. This shows that not only can the spectrum of quantum light be manipulated by application of a well-controlled temporal phase, but this method also preserves the spectral phase of the light and thus the wave-packet pulse mode. 

The spectral shear operation based upon electro-optic temporal phase modulation presented here demonstrates a number of essential features required for photonic TF encoding of quantum light. The integrated-optics platform upon which this method is based is well suited to existing and developing integrated quantum photonics efforts, where small footprint, low loss, and high stability are essential \cite{smith09,obrien13}. The phase-only, linear optical nature of the process does not introduce photon-number noise and ensures deterministic operation. We have shown by two-photon interference that there is no decoherence of single-photon wave packets. Indeed, the ability to fine-tune the two-photon interference by shifting the central frequency of single-photon wave packets demonstrated here could be used to match single-photon emitters, such as quantum dots \cite{buckley12}. Loss in the system arises primarily from coupling in and out of the modulator waveguide, which can, in principle, be addressed. The application of a well-defined temporal phase by electro-optic phase modulation to quantum pulses is readily tunable and can be used to implement a variety of operations that, when paired with a controlled spectral phase, forms the basis of a quantum pulse shaper \cite{brecht11,kielpinski11,morizur10}. This work demonstrates a key principle of general arbitrary phase modulation, the application of linear temporal phase, and paves the way for arbitrary pulse shaping of broadband quantum light pulses. It also provides demonstration of a valuable interface between optical and rf signals that may be utilized for control and measurement in the quantum domain. This technique will likely play a central role in the emergent field of optical quantum technologies, where temporal-spectral multiplexing enables significant advantages for integrated optics.

We are grateful to K.\ Banaszek, M.\ Cooper, C.\ Dorrer, D.\ Oblak, D.\ Gauthier, C.\ Radzewicz, M.~G.\ Raymer, N.\ Sinclair, and I.~A.\ Walmsley for fruitful discussions. We thank Aspen Electronics for assistance in designing and assembling the rf electronics and Justin Spring for assistance programming the FPGA. This project has received funding from the European Union's Horizon 2020 research and innovation programme under Grant Agreement No.\ 665148. M.~K. was partially supported by a Marie Curie Intra-European Fellowship No.\ 301032 and by the PhoQuS@UW project (Grant Agreement No.\ 316244) within the European Community 7th Framework Programme, as well as by the National Science Centre of Poland project No.\ 2014/15/D/ST2/02385.

\medskip

{\em Note added in the proof.}---Recently, we became aware of related work by Fan {\em et el.} \cite{fan2016}.

\end{document}